\documentclass[twocolumn,showpacs,aps,prd,superscriptaddress,letterpaper]{revtex4}

\usepackage{graphicx}
\usepackage{dcolumn}
\usepackage{amsmath}
\usepackage{epsfig}

\newcommand{\acp}{\ensuremath{\mathcal A_{CP}}\xspace}

\long\def\inst#1{\par\nobreak\kern 4pt\nobreak
    {\itshape #1}\par\vskip 10pt plus 3pt minus 3pt}

\def\btosss{\ensuremath{b\to\s\bar{s}s}\xspace}

\RequirePackage{xspace}

\usepackage{relsize}
\def\babar{\mbox{\slshape B\kern-0.37em{\relsize{-2} A}\kern-0.04em
    B\kern-0.37em{\relsize{-2} A\kern-0.04em R}}\xspace}

\def\epem       {\ensuremath{e^+e^-}\xspace}

\def\s     {\ensuremath{s}\xspace}

\def\pip   {\ensuremath{\pi^+}\xspace}
\def\pim   {\ensuremath{\pi^-}\xspace}

\def\pipm  {\ensuremath{\pi^\pm}\xspace}

\def\Kbar  {\kern 0.2em\overline{\kern -0.2em K}{}\xspace}

\def\Kz    {\ensuremath{K^0}\xspace}
\def\Kzb   {\ensuremath{\Kbar^0}\xspace}
\def\KzKzb {\ensuremath{\Kz \kern -0.16em \Kzb}\xspace}
\def\Kp    {\ensuremath{K^+}\xspace}
\def\Km    {\ensuremath{K^-}\xspace}
\def\Kpm   {\ensuremath{K^\pm}\xspace}

\def\KpKm  {\ensuremath{\Kp \kern -0.16em \Km}\xspace}
\def\KS    {\ensuremath{K^0_{\scriptscriptstyle S}}\xspace}

\def\Dbar    {\kern 0.2em\overline{\kern -0.2em D}{}\xspace}

\def\Dz      {\ensuremath{D^0}\xspace}
\def\Dzb     {\ensuremath{\Dbar^0}\xspace}
\def\DzDzb   {\ensuremath{\Dz {\kern -0.16em \Dzb}}\xspace}
\def\Dp      {\ensuremath{D^+}\xspace}
\def\Dm      {\ensuremath{D^-}\xspace}

\def\DpDm    {\ensuremath{\Dp {\kern -0.16em \Dm}}\xspace}

\def\Dstarp  {\ensuremath{D^{*+}}\xspace}

\def\Bbar    {\kern 0.18em\overline{\kern -0.18em B}{}\xspace}

\def\BB      {\ensuremath{B\Bbar}\xspace} 
\def\Bz      {\ensuremath{B^0}\xspace}
\def\Bzb     {\ensuremath{\Bbar^0}\xspace}
\def\BzBzb   {\ensuremath{\Bz {\kern -0.16em \Bzb}}\xspace}
\def\Bu      {\ensuremath{B^+}\xspace}
\def\Bub     {\ensuremath{B^-}\xspace}
\def\Bp      {\ensuremath{\Bu}\xspace}
\def\Bm      {\ensuremath{\Bub}\xspace}
\def\Bpm     {\ensuremath{B^\pm}\xspace}

\def\BpBm    {\ensuremath{\Bu {\kern -0.16em \Bub}}\xspace}

\newcommand{\optbar}[1]{  \shortstack[r]{ \\[-1.0ex] { \tiny (\rule[.4ex]{.8em}{.1mm}) } \\ [-1.0ex] $#1$}} 
\def\BorBbar    {\kern 0.18em\optbar{\kern -0.18em \Bz}{}\xspace}
\def\DorDbar    {\kern 0.18em\optbar{\kern -0.18em D}{}\xspace}
\def\KorKbar    {\kern 0.18em\optbar{\kern -0.18em \Kz}{}\xspace}

\mathchardef\Upsilon="7107
\def\Y#1S{\ensuremath{\Upsilon{(#1S)}}\xspace}

\def\FourS {\Y4S}

\mathchardef\Deltares="7101
\mathchardef\Xi="7104
\mathchardef\Lambda="7103
\mathchardef\Sigma="7106
\mathchardef\Omega="710A

\def\Deltabar{\kern 0.25em\overline{\kern -0.25em \Deltares}{}\xspace}
\def\Lbar{\kern 0.2em\overline{\kern -0.2em\Lambda\kern 0.05em}\kern-0.05em{}\xspace}
\def\Sigbar{\kern 0.2em\overline{\kern -0.2em \Sigma}{}\xspace}
\def\Xibar{\kern 0.2em\overline{\kern -0.2em \Xi}{}\xspace}
\def\Obar{\kern 0.2em\overline{\kern -0.2em \Omega}{}\xspace}
\def\Nbar{\kern 0.2em\overline{\kern -0.2em N}{}\xspace}
\def\Xb{\kern 0.2em\overline{\kern -0.2em X}{}\xspace}

\def\BR         {{\ensuremath{\mathcal B}\xspace}}

\def\pt         {\ensuremath{p_T}\xspace}
\def\mes        {\ensuremath{m_{\text{ES}}}\xspace}

\def\DeltaE     {\ensuremath{\Delta E}\xspace}

\newcommand{\tev}{\ensuremath{\mathrm{\,Te\kern -0.1em V}}\xspace}
\newcommand{\gev}{\ensuremath{\mathrm{\,Ge\kern -0.1em V}}\xspace}
\newcommand{\mev}{\ensuremath{\mathrm{\,Me\kern -0.1em V}}\xspace}
\newcommand{\kev}{\ensuremath{\mathrm{\,ke\kern -0.1em V}}\xspace}
\newcommand{\ev}{\ensuremath{\mathrm{\,e\kern -0.1em V}}\xspace}
\newcommand{\gevc}{\ensuremath{{\mathrm{\,Ge\kern -0.1em V\!/}c}}\xspace}
\newcommand{\mevc}{\ensuremath{{\mathrm{\,Me\kern -0.1em V\!/}c}}\xspace}
\newcommand{\gevcc}{\ensuremath{{\mathrm{\,Ge\kern -0.1em V\!/}c^2}}\xspace}
\newcommand{\mevcc}{\ensuremath{{\mathrm{\,Me\kern -0.1em V\!/}c^2}}\xspace}

\def\to                 {\ensuremath{\rightarrow}\xspace}

\def\pep2{PEP-II}

\newcommand{\dedx}{\ensuremath{\mathrm{d}\hspace{-0.1em}E/\mathrm{d}x}\xspace}

\def\gsim{{~\raise.15em\hbox{$>$}\kern-.85em
          \lower.35em\hbox{$\sim$}~}\xspace}
\def\lsim{{~\raise.15em\hbox{$<$}\kern-.85em
          \lower.35em\hbox{$\sim$}~}\xspace}

\def\eps{\varepsilon\xspace}

\def\CP                {\ensuremath{C\!P}\xspace}

\xspace

\newcommand{\jprlBase}       {Phys.\ Rev.\ Lett.\xspace}
\newcommand{\jprBase}        {Phys.\ Rev.\xspace}
\newcommand{\jplBase}        {Phys.\ Lett.\xspace}

\newcommand{\nimBaseE}       {Nucl.\ Instrum.\ Methods\ Phys.\ Res.\xspace}

\newcommand{\zpBase}         {Z.\ Phys.\xspace}

\newcommand{\nima}      [1]  {\nimBaseE~A~{\bfseries #1}}

\newcommand{\plb}       [1]  {\jplBase\ B~{\bfseries #1}}

\newcommand{\jprl}      [1]  {\jprlBase\ {\bfseries #1}}
\newcommand{\jprd}      [1]  {\jprBase\ D~{\bfseries #1}}

\newcommand{\progtp}    [1]  {{Prog.\ Theor.\ Phys.\ {\bfseries #1}}}

\newcommand{\zpc}       [1]  {\zpBase\ C~{\bfseries #1}}

\def\jetset74   {\text{\ttfamily Jetset \hspace{-0.5em}7.\hspace{-0.2em}4}\xspace}

\newcommand{\BABARPubYear}    {03}
\newcommand{\BABARPubNumber}  {17}

\newcommand{\SLACPubNumber} {10086}

\begin{document}

\begin{flushleft}
\babar-PUB-\BABARPubYear/\BABARPubNumber\\
SLAC-PUB-\SLACPubNumber\\
[10mm]
\end{flushleft}


\title{
%
\large \bfseries Measurements of branching fractions
in \boldmath{$B \to \phi K$} and \boldmath{$B \to \phi \pi$} \\
and search for direct \CP violation in {$\Bpm \to \phi \Kpm$}
}

%
\author{B.~Aubert}
\author{R.~Barate}
\author{D.~Boutigny}
\author{J.-M.~Gaillard}
\author{A.~Hicheur}
\author{Y.~Karyotakis}
\author{J.~P.~Lees}
\author{P.~Robbe}
\author{V.~Tisserand}
\author{A.~Zghiche}
\affiliation{Laboratoire de Physique des Particules, F-74941 Annecy-le-Vieux, France }
% ...
\collaboration{\babar\ Collaboration}
\noaffiliation


\date{January 29, 2004}


\begin{abstract}
We present measurements of branching 
fractions in the \btosss penguin-dominated
decays $\Bp\to \phi \Kp$ and $\Bz\to \phi \Kz$
in a sample of approximately 89 million \BB pairs
collected by the \babar detector at the \pep2\ asymmetric-energy \emph{B}-meson 
factory at SLAC.
We determine ${\mathcal B}(B^+ \to \phi K^+) = 
(10.0^{+0.9}_{-0.8} \pm 0.5) \times 10^{-6}$ and 
${\mathcal B}(B^0 \to \phi K^0) = (8.4^{+1.5}_{-1.3} \pm 0.5) \times 10^{-6}$.
Additionally, we measure the \CP-violating charge asymmetry $\acp (B^{\pm} \to \phi K^{\pm}) =
0.04 \pm 0.09 \pm 0.01$, with a 90\% confidence-level interval of $\left [-0.10,0.18 \right ]$,
and set an upper limit on the CKM-- and color-suppressed
decay $B^+ \to \phi \pi^+$,
${\mathcal B}(B^+ \to \phi \pi^+) < 0.41 \times 10^{-6}$ (at the 90\% confidence level).
\end{abstract}


\maketitle

\begin{figure}[hpt]
\epsfig{file=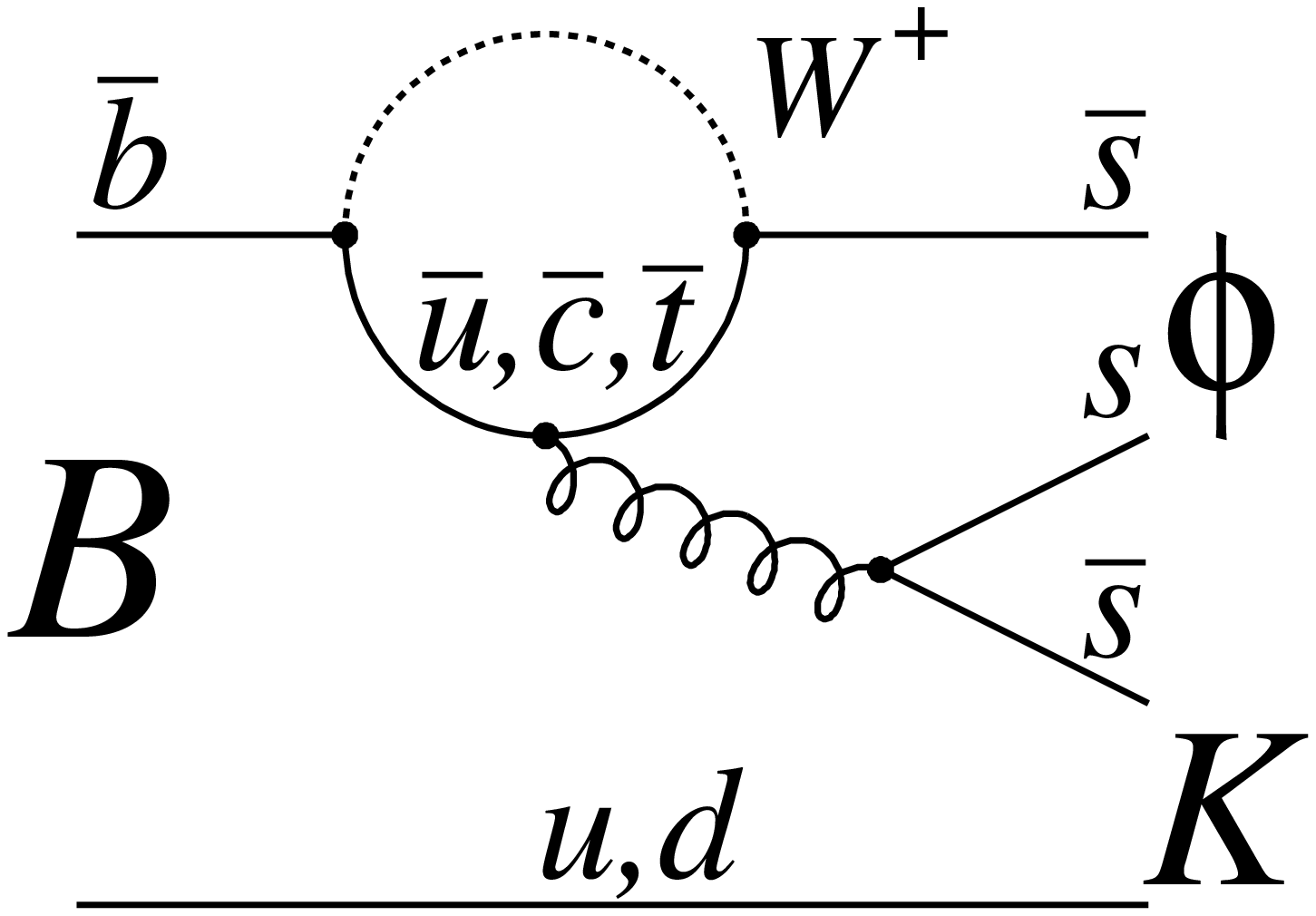, width=4.25cm} 
\epsfig{file=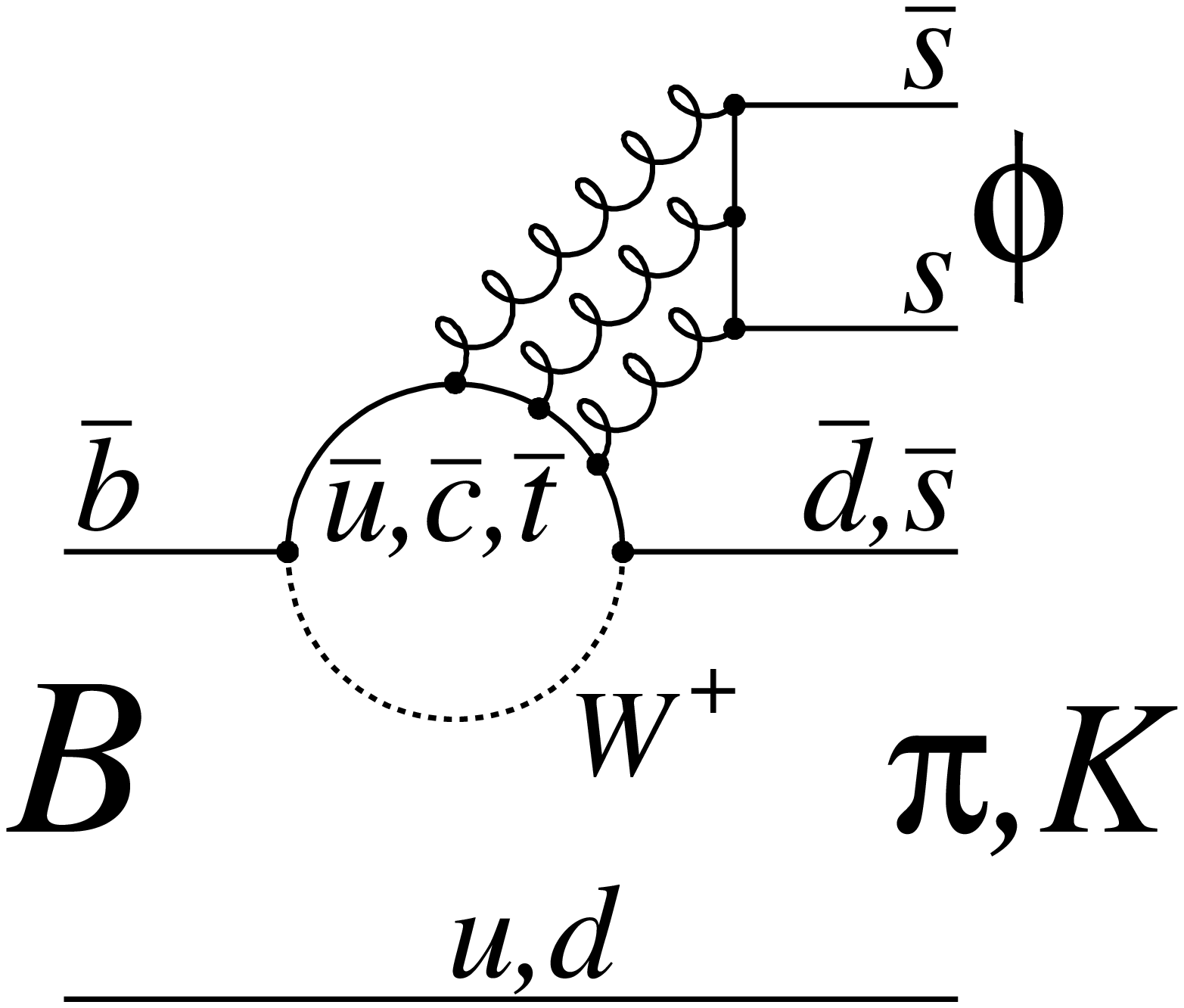, width=4.25cm} 
\caption{Examples of quark-level diagrams for $B\to\phi K$ 
and $B\to\phi \pi$. Left: internal penguin diagram, 
right: flavor-singlet penguin diagram.\label{fig:diagram}}
\end{figure}

Decays of $B$ mesons into charmless hadronic final states with a 
$\phi$ meson are dominated by $b\to s\bar{s}s$ gluonic penguin diagrams 
(Fig.~\ref{fig:diagram}), possibly with 
smaller contributions from electroweak penguin diagrams, while other
Standard Model (SM) amplitudes are strongly suppressed~\cite{one}.
In the Standard Model,  \CP violation arises from a single complex phase in the
Cabibbo--Kobayashi--Maskawa (CKM) quark-mixing matrix~\cite{ckm}.
Since many scenarios of physics beyond the SM introduce additional diagrams 
with heavy particles in the penguin loops and new \CP-violating 
phases~\cite{grossman}, a comparison of \CP-violating observables with SM 
expectations is a sensitive probe for new physics. 
In the SM, neglecting CKM-suppressed contributions, 
the direct \CP violation in $B^+ \to \phi K^+$~\cite{charge}, 
detected as an asymmetry 
$\acp = (\Gamma_{\phi K^-} - \Gamma_{\phi K^+})/(\Gamma_{\phi K^-} + \Gamma_{\phi K^+})$ 
in the decay rates $\Gamma_{\phi K^\pm} = \Gamma(B^\pm \to \phi K^\pm)$, 
is expected to be zero; in the presence of large new-physics contributions to
the $b\to s\bar{s}s$ transition, it could be of order 1~\cite{Ciuchini:2002pd}.
The $B \to \phi K$ and $B \to \phi \pi$ decay rates are also 
sensitive to new physics; the latter is strongly suppressed in the SM, 
and a measurement of ${\mathcal B}(B \to \phi \pi) \gtrsim 10^{-7}$
would serve as evidence for new physics~\cite{Bar-Shalom:2002sv}.
The branching fractions of $B^+ \to \phi K^+$  and $\Bz \to \phi \Kz$ 
have been studied by CLEO~\cite{Briere:2001ue}, 
\babar~\cite{oldpub,prd}, and Belle~\cite{:2003jf}; 
$\acp (B^+ \to \phi K^+)$ has been studied by \babar~\cite{prd}.

This analysis is based on an integrated luminosity of about 82~fb$^{-1}$, corresponding 
to approximately 89 million \BB pairs, collected 
at SLAC with the \babar detector~\cite{Aubert:2001tu} at the \pep2\ 
asymmetric-energy \epem storage ring operating on the $\Upsilon(4S)$ 
resonance. 

The asymmetric beam configuration
provides a boost to the \FourS in the laboratory frame ($\beta\gamma\approx 0.56$),
increasing the maximum momentum of the $B$-meson decay products
to $4.4 \gevc$.
Charged particles are detected and their momenta measured
by a combination of a silicon vertex tracker (SVT), consisting 
of five double-sided layers, and a 40-layer central drift chamber (DCH), 
both operating in a 1.5~T solenoidal magnetic field. 
The tracking system covers 92\% of the solid angle
in the center-of-mass (CM) frame.
The track-finding efficiency is, on average, ($98\pm1$)\% for momenta
above $0.2 \gevc$ and polar angles greater than 0.5~rad. 
Photons are detected by a CsI(Tl) electromagnetic calorimeter (EMC), which
provides excellent angular and energy resolution with high efficiency for 
energies above 20~\mev.

Charged-particle identification is provided by measuring the average 
energy loss (\dedx) in the two tracking devices and
by the novel internally reflecting ring-imaging 
Cherenkov detector (DIRC) covering the central region. 
A $\pi/K$ separation of better than $4\sigma$ is 
achieved for tracks with momenta below $3 \gevc$, decreasing to 
$2.4\sigma$ for the highest momenta arising from $\Bu \to \phi h^+$ decays. 
Electrons are identified with the use
of the tracking system and the EMC.


We fully reconstruct $B$-meson candidates in the decay modes 
$\phi h^+$ and $\phi \KS$, with $\phi\rightarrow K^+K^-$ and
$\KS\rightarrow\pi^+\pi^-$.
For the $h^+$ track and the charged-track daughters of the $\phi$
we require at least 12 measured 
DCH hits and a minimal transverse momentum \pt of 0.1~\gevc.
The tracks must originate from the interaction point 
(within 10~cm along the beam direction and 1.5~cm in the transverse plane).
Looser criteria are applied to tracks belonging to
$\KS\to\pi^+\pi^-$.
We combine pairs of oppositely charged tracks 
originating from a common vertex to form \KS and $\phi$ candidates.
A $\KS\rightarrow\pi^+\pi^-$ candidate is accepted on 
the basis of requirements on the two-pion invariant mass
(within $12$~\mevcc of the nominal \KS\ mass~\cite{pdg}), 
the flight-length ($\ell$) significance ($\ell/\sigma_{\ell} >3$), 
and the angle between the line connecting the $B$ and \KS\ decay vertices
and the \KS momentum ($<0.1$~rad).
Kaon tracks used to reconstruct the $\phi$ meson are distinguished from pion and proton tracks
using \dedx information from the DCH in conjunction with \dedx information from 
the SVT
for track momenta below 0.7 \gevc, and, for momenta above 0.7 \gevc, with the
measured Cherenkov angle and number of photons recorded by the 
DIRC.

For an extended unbinned maximum-likelihood (ML) fit we parameterize the 
distributions of kinematic and topological variables for signal and 
background events in terms of probability density functions (PDFs).
Each $B$ candidate is characterized by the energy 
difference $\Delta E = (q_{\Upsilon} \cdot q_{B}/\sqrt{s}) - \sqrt{s}/2$ and
the beam-energy--substituted mass
$\mes = [({s}/{2} + {\vec{p}}_{\Upsilon} \cdot 
{\vec{p}}_{B} )^{2}/{E^{2}_{\Upsilon}} - 
{{\vec{p}}_{B}}^{\,2}]^{1/2}$~\cite{Aubert:2001tu}.
Here $q_{\Upsilon}$\/ and $q_{B}$ are four-momenta of the \Y4S and the $B$\/ 
candidate, $s \equiv (q_{\Upsilon})^{2}$ is the square of the 
center-of-mass energy, ${\vec{p}}_{\Upsilon}$ 
and ${\vec{p}}_{B}$ are the three-momenta of the \Y4S\/ and the $B$\/ in the laboratory frame, 
and ${E_{\Upsilon}} \equiv q^0_{\Upsilon}$ is the energy of the \Y4S in the laboratory frame.
For signal events, \DeltaE peaks at zero and \mes peaks at the nominal $B$ mass. 
The signal PDFs of both variables are adequately described by sums of two Gaussian distributions
(whose means are not required to be the same).
The background shape in $\Delta E$ is parametrized by a linear function
and in $m_{ES}$ by a threshold function~\cite{argus}.
Candidates for our analysis are required to satisfy $|\DeltaE|<0.2 \gev$ and $\mes > 5.2 \gevcc$. 
The variable \DeltaE provides additional momentum-dependent $\pi/K$
separation in the ML fit for the $\Bp \to \phi h^+$ branching fractions.
The likelihood also incorporates the invariant mass of the $\phi \to \Kp \Km$ candidate
$m_{KK}$ in the $[0.99, 1.05]$ \gevcc range, which is described by a 
relativistic Breit--Wigner function 
convolved with a Gaussian, $\sigma = 1.0 \mevcc$, determined in Monte Carlo (MC) simulation studies,
to account for resolution 
effects, and the $\phi$ helicity angle $\theta_H$, which is defined 
as the angle between the directions of the $K^+$ and the parent $B$
in the $\phi$ rest frame. The $\cos\theta_H$ distribution is a quadratic
function for pseudoscalar-vector $B$ decay modes and is nearly uniform 
for the combinatorial background.

Backgrounds in the candidate sample arise primarily from random combinations of tracks 
produced in the quark-antiquark continuum. 
In such events, particles appear bundled into jets, which can be
identified with several variables computed in the CM frame.
We use the angle $\theta_T$ between the thrust axis of the $B$ candidate 
and the thrust axis of the other charged and neutral particles~\cite{Aubert:2001tu}. 
We require the angle $\theta_T$ to satisfy $|\cos\theta_T| < 0.9$.
Other quantities that characterize the event topology are 
the CM angle $\theta_B$ between the $B$ momentum and the beam axis
and the sum of the momenta $p_i$ of the other 
charged and neutral particles
in the event weighted with Legendre polynomials $L_n(\theta_i), n=0,2$,
where $\theta_i$ is the angle between the momentum of particle $i$
and the thrust axis of the $B$ candidate.
We combine these variables 
into a Fisher discriminant ${\mathcal F}$~\cite{Fisher:et}.
Contamination from other $B$ decays, as well as 
$\tau^+\tau^-$ and $e^+e^-\gamma\gamma$ production, is negligible,
as demonstrated in 
MC simulation studies.
Possible $K^+K^-$ $S$-wave contributions, such as the $f_0(980)$ and the $a_0(980)$, 
are not
expected to contribute under the $\phi$ mass peak~\cite{scalar} and are
distinguished by their uniform distribution in $\cos\theta_H$;
this systematic effect is small compared with current statistical and
systematic uncertainties.

We use an unbinned extended ML fit to extract
signal yields and charge asymmetries simultaneously.
The likelihood for candidate $j$ in the flavor category $c$ 
is obtained by summing the product of event yield $N_{ic}$ and probability 
${\mathcal P}_{ic}$ over signal and background hypotheses $i$.
The total extended likelihood ${\mathcal L}$ for a sample of $N$ events is given by 
\begin{equation}
{\mathcal L} = \frac{1}{N!}\exp{\left(-\sum_{i,c}N_{ic}\right)}
\prod_{j=1}^N\left[\sum_{i,c}N_{ic}{\mathcal P}_{ic}(\vec{x}_j;\vec{\alpha}_i)\right]\!.
\end{equation}
The probabilities ${\mathcal P}_{ic}$ 
are products of PDFs for each of the independent variables 
$\vec{x}_j = \left\{\mes, \DeltaE, {\mathcal F}, m_{KK}, \cos\theta_H \right\}$. 
The $\vec{\alpha}_i$ are the parameters of the distributions in $\vec{x}_j$,
which are fixed to values derived from signal MC,
on-resonance sidebands in (\mes, \DeltaE), and high-statistics data control channels
$\Bu \to \pip \Dzb \ (\Dzb \to \Kp \pim)$
and $\Bz \to \pip \Dm \ (\Dm \to \KS \pim)$. The control channels have event topologies similar
to those in $\Bu \to \phi \Kp$ and $\Bz \to \phi\KS$, and
are used to compare central values and resolutions of the variables \mes, \DeltaE, and ${\mathcal F}$ in 
data and MC simulation.
By minimizing the quantity $-\ln{\mathcal L}$ in two separate fits, 
we determine the branching fractions, \BR, and the charge asymmetry, \acp, for $\phi h^\pm$ and 
$\phi \KS$. In the $\phi\KS$ case, there are two hypotheses, 
signal and background ($i=1,2$), and a single flavor category.
In the 
fit for $\Bpm \to \phi h^\pm$ decays, we determine the flavor of the high-momentum track by 
comparing the measured Cherenkov angle with that expected for a pion or a kaon.
In this way, the $\phi h^\pm$ ($h=\pi, K$) decays
are fitted simultaneously with two signal 
($i=1$ for $\Bpm \to \phi \Kpm$ and 
$i=2$ for $\Bpm \to \phi \pipm$)
and two corresponding background ($i=3,4$) hypotheses. 
We define the event yields $n_{ic}$ in each of the two flavor
categories ($c = 1$ for $\Bp \to \phi h^+$ and 
$c = 2$ for $\Bm \to \phi h^-$) in terms of 
the charge asymmetry ${\cal A}_i$ and the total event yield $n_{i}$:
$n_{i1} = n_{i}\times(1 + {\cal A}_i)/2$ and
$n_{i2} = n_{i}\times(1 - {\cal A}_i)/2$.

\begingroup
\begin{table}[bt]
\caption{\label{tb:results}Summary of branching fraction (\BR) and direct \CP-asymmetry (\acp) results.
$N_{\text{sig}}$ and $\eps$ are the signal yield and the total efficiency in the 
branching fraction fit.
The 90\% confidence-level interval for \acp\ is $[-0.10, 0.18]$.
}
\begin{ruledtabular}
\setlength{\extrarowheight}{1.5pt}
\footnotesize
\begin{tabular}{lcccc}
Mode            & $\eps$ (\%) & $N_{\text{sig}}$  &     \BR\ ($10^{-6}$)         &      \acp \\
\hline
$\phi K^0$      & $6.7$ & ${50^{+9}_{-8}}$      & $8.4^{+1.5}_{-1.3}\pm0.5$ & 
 --- \\
\hline
$\phi K^+$    & 19.6 & $173\pm15$          & $10.0^{+0.9}_{-0.8}\pm0.5$     &  $0.04 \pm 0.09 \pm 0.01$\\
\hline
$\phi \pi^+$  & 20.4 & $0.9^{+2.4}_{-0.9}$ & $< 0.41$~(90\% CL)         & --- \\
\end{tabular}
\setlength{\extrarowheight}{0pt}
\end{ruledtabular}
\end{table}
\endgroup

For charged tracks originating from the interaction point, we determine the ratio of 
track-finding efficiencies in data and MC simulation by 
conducting a study of a large sample of unambiguous charged-track candidates that
have at least 10 measured hits in the SVT; the method relies on the fact that
for both the SVT and the DCH the
differences between the track-finding efficiencies in data and MC simulation 
are small, and so the two detectors can be used to calibrate each other.
The ratio of $\KS \to \pip \pim$ reconstruction efficiencies in data
and MC simulation as a function of the \KS momentum and decay point is determined from
a study of a large inclusive sample of $\KS \to \pip \pim$  decays; this method employs the 
results of the tracking-efficiency study that covers \KS decays occuring in the immediate vicinity of
the interaction point.
 The charged-kaon--identification
efficiencies in data and MC simulation are compared in a study of fully reconstructed 
$\Dstarp \to \Dz\pip 
(\Dz \to \Km \pip)$ decays.

\begin{figure}[t]
\center
\epsfig{file=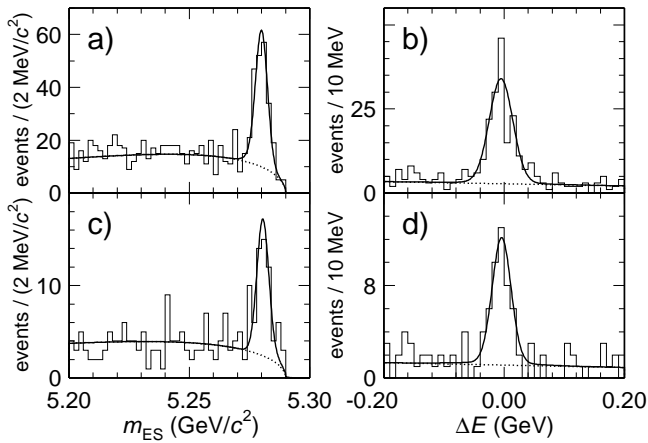, width=8.5cm}
\caption{Projection plots of the variables \mes [(a) and (c)] 
and \DeltaE [(b) and (d)] in the fit
for the $\phi K^+$ (top) and $\phi \KS (\pi^+\pi^-)$ (bottom)
branching fractions.
The data are shown by the histogram, while the curve is the result of the
fit.
The signal-to-background ratio is enhanced with a requirement on the signal
probability ${\cal P}_{\mathrm{sig}}/({\cal P}_{\mathrm{sig}}+{\cal P}_{\mathrm{bkg}})$
with the PDF for the variable being plotted excluded.
\label{fig:yield}}
\end{figure}

Results of the branching-fraction and \CP-asymmetry fits are given in 
Table~\ref{tb:results}. 
Equal production rates of \BzBzb and \BpBm are assumed.
Figure~\ref{fig:yield} shows the \mes and \DeltaE distributions 
of $\phi\KS (\pi^+\pi^-)$ and $\phi K^+$ events together with the likelihood 
projections from the \BR\ fits.
Goodness-of-fit tests have been performed to confirm that the values
of likelihood $\mathcal L$ obtained in the fits are consistent with MC-based
expectations.

Systematic uncertainties in the ML fit originate from assumptions about
the signal and background distributions and are dominated by the limited 
sideband and control-channel statistics.
We simultaneously vary all PDF parameters within their uncertainties,
and derive the associated systematic errors: 0.005 for \acp, 
2.0\% for $\BR(\phi\Kp)$,
and 2.8\% for $\BR(\phi K^0)$.
To account for the systematic uncertainty on the upper limit
on $\BR(\phi\pip)$, we increase the upper limit
by one standard deviation due to PDF variations (10.9\%) and due
to uncertainty in the reconstruction efficiency (4.2\%).
The dominant systematic errors in the efficiency come from track 
finding (2.4\% for $\BR(\phi h^+)$ and 4.2\% for $\BR(\phi\KS)$),
charged-kaon identification (2\% per $\phi$), 
and \KS
reconstruction efficiency (2\%). 
Other systematic errors from event-selection criteria, daughter branching fractions,
MC statistics, \BB backgrounds and $B$-meson counting sum in 
quadrature to 3.0\%.
The systematic uncertainty
on \acp due to charge asymmetries in tracking and the DIRC is less than 0.01.

In summary, we have studied branching fractions and charge asymmetries
in the $B$-meson final states $\phi h^+$ and $\phi \KS$; the
results are listed in Table~\ref{tb:results}.
We do not observe a significant charge asymmetry in the mode 
$\Bp\to \phi \Kp$ and do not see evidence for $B^+ \to \phi \pi^+$.
Our branching fraction and charge asymmetry measurements are consistent with, 
and supersede, our previous results reported in~\cite{oldpub,prd}.
They are also consistent with existing SM predictions.

We are grateful for the excellent luminosity and machine conditions
provided by our \pep2\ colleagues, 
and for the substantial dedicated effort from
the computing organizations that support \babar.
The collaborating institutions wish to thank 
SLAC for its support and kind hospitality. 
This work is supported by
DOE
and NSF (USA),
NSERC (Canada),
IHEP (China),
CEA and
CNRS-IN2P3
(France),
BMBF and DFG
(Germany),
INFN (Italy),
FOM (The Netherlands),
NFR (Norway),
MIST (Russia), and
PPARC (United Kingdom). 
Individuals have received support from the 
A.~P.~Sloan Foundation, 
Research Corporation,
and Alexander von Humboldt Foundation.

\bibliographystyle{h-physrev2-original}

\end{document}